\documentclass{elsart}

\usepackage{graphicx}

\usepackage{amssymb}

\begin{document}

\def\cH{{\mathcal H}}
\def\cS{{\mathcal S}}

\newcommand{\beq}{\begin{equation}}
\newcommand{\eeq}{\end{equation}}
\newcommand{\barr}{\begin{eqnarray}}
\newcommand{\earr}{\end{eqnarray}}

\begin{frontmatter}



\title{Hausdorff clustering of financial time series}


\author[1]{Nicolas Basalto},
\ead{nicolas.basalto@unipv.it}
\author[2,3,4]{Roberto Bellotti},
\author[2,3]{Francesco De Carlo},
\author[2,3]{Paolo Facchi},
\author[2,3]{Saverio Pascazio}

\address[1]{Institute for
Advanced Studies at University of Pavia, Italy.}
\address[2]{Dipartimento di Fisica, Universit\`a di Bari, Italy.}
\address[3]{Istituto Nazionale di Fisica Nucleare, Sezione di Bari,
Italy.}
\address[4]{TIRES, Center of Innovative Technologies for
Signal Detection and Processing, Bari, Italy.}

\begin{abstract}
A clustering procedure, based on the Hausdorff distance, is
introduced and tested on the financial time series of the Dow
Jones Industrial Average (DJIA) index.
\end{abstract}

\end{frontmatter}

\section{Introduction}
\label{sec:intro}
Clustering consists in grouping a set of objects in classes
according to their degree of ``similarity" \cite{fukunaga}. This
intuitive concept can be defined in a number of different ways,
leading in general to different partitions. For this reason, it is
clear that a clustering procedure can be profoundly influenced by
the strategy adopted by the observer and his/her own ideas and
preconceptions about the data set. In this article we will focus
on a \emph{linkage} algorithm, that consists in merging, at each
step, the two clusters with the smallest dissimilarity, starting
from clusters made up of a single element and ending up in a
single cluster collecting all data. Our objective will be to
cluster the financial time series of the stocks belonging to the
Dow Jones Industrial Average (DJIA) index.

From a mathematical point of view, given a set of objects
$\mathcal{S}\equiv\{ s \}$, an allocation function
$m:\mathcal{S}\rightarrow \{1,2,\dots,k\}$, is defined so that
$m(s)$ is the class label and \textit{k} the total number of
clusters (which we assume to be finite for simplicity). The aim of
a clustering procedure is to select, among all possible allocation
functions, the one performing the best partition of the set $\cS$
into subsets $\mathcal{G}_{\alpha}\equiv\{s \in \mathcal{S} | m(s)
= \alpha\}, (\alpha=1,\dots,k )$, relying on some measure of
similarity.

Clustering algorithms can be classified in different ways
according to the criteria used to implement them. The so-called
``hierarchical" methods yield nested partitions, represented by
\textit{dendrograms} \cite{jain}, in which any cluster can be
further divided in order to observe its underlying structure.
Linkage algorithms, in particular, are hierarchical. Other
non-hierarchical (or ``partitional") methods are also possible
\cite{central,duda,hofmann}, but will not be discussed here.

\section{Hausdorff clustering}
\label{sec:Hausdorff}

In order to cluster a given data set we will use a distance
function introduced by Hausdorff. Given a metric space $(\cS,
\delta)$, with metric $\delta$, the distance between a point
$a\in\cS$ and a subset $B \subseteq \cS$ is naturally given by
\begin{equation}
\tilde d(a; B)=\inf_{b\in B} \delta(a,b)
\end{equation}
(all subsets are henceforth considered to be non-empty and
compact). Given a subset $A \subseteq \cS$, let us define the
function
\beq
\tilde d(A;B)=\sup_{a\in A} \tilde d(a; B) = \sup_{a\in
A}\inf_{b\in B} \delta(a,b),
\label{openhaus}
\eeq
which measures the largest among all distances $\tilde d(a; B)$,
with $a\in A$. This function is not symmetric, $\tilde d(A;B)
\neq\tilde d(B;A)$, and therefore is not a \emph{bona fide}
distance. The Hausdorff distance \cite{edgar} between two sets
$A,B \subseteq \cS$ is defined as the largest between the two
numbers:
\barr
d_{\rm H}(A,B)&=&\max\{\tilde d(A;B), \tilde d(B;A)\}
\nonumber \\
& =& {\max}\{\sup_{a\in A}\inf_{b\in B}\delta(a,b), \sup_{b\in
B}\inf_{a\in A} \delta(a,b) \}
\label{link_haus}
\earr
and is clearly symmetric.

\begin{figure}
\begin{center}
\includegraphics[width=0.5\textwidth]{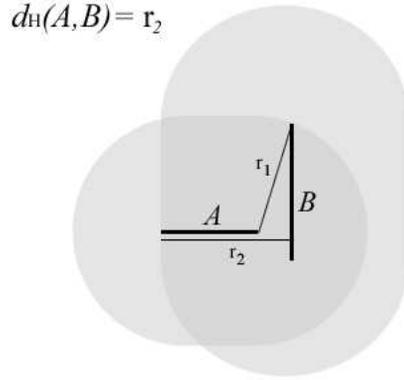}
\end{center}
\caption{Hausdorff distance between two sets $A$ and $B$ (black
thick segments). $r_1=\tilde d(B;A)$, $r_2=\tilde d(A;B)$. The
Hausdorff distance is equal to the larger radius $r_{2}$.}
\label{fig:dist_H_AB}
\end{figure}

In words, the Hausdorff distance between $A$ and $B$ is the
smallest positive number $r$, such that every point of $A$ is
within distance $r$ of some point of $B$, \emph{and} every point
of $B$ is within distance $r$ of some point of $A$. The meaning of
the Hausdorff distance is best understood by looking at an
example, such as that in Fig.\ \ref{fig:dist_H_AB}. We emphasize
that the Hausdorff metric relies on the metric $\delta$ on $\cS$.

If the data set is finite and consists of $N$ elements, all
distances can be arranged in a $N \times N$ matrix $\delta_{ij}$
and Eq.\ (\ref{link_haus}) reads
\beq
d_{\rm H}(A,B)=\max\{ \max_{i\in A}\min_{j\in B} \delta_{ij}~,
\max_{j\in B}\min_{i\in A} \delta_{ij}\},
\label{link_haus_finite}
\eeq
which is a very handy expression, as it amounts to finding the
minimum distance in each row (column) of the distance matrix, then
the maximum among the minima. The two numbers are finally compared
and the largest one is the Hausdorff distance. This sorting
algorithm is easily implemented in a computer.

The Hausdorff distance naturally translates in a linkage
algorithm. At the first level each element is a cluster and the
Hausdorff distance between any pair of points reads
\beq
d_{\rm H}(\{i\},\{j\}) = \delta_{ij}
\eeq
and coincides with the underlying metric.

The two elements of $\cS$ at the shortest distance are joined
together in a single cluster. The Hausdorff distance matrix is
recomputed, considering the two joined elements as a single set.
This iterative process goes on until all points belong to a single
final cluster.

\section{Comparison with single and complete linkage}
\label{sec:compSC}

It is interesting to notice that the partitions obtained by the
Haudorff linkage algorithm are intermediate between those obtained
by the more commonly used ``single" and ``complete" linkage
procedures: if $A$ and $B$ are two non empty subsets of $\cS$, the
single and complete linkage algorithms make use of the following
similarity indexes
\barr
d_{\rm S}(A,B) &=& \inf_{a\in A, b\in B} \delta(a,b), \label{dist_min} \\
d_{\rm C}(A,B) &=& \sup_{a\in A, b\in B} \delta(a,b) ,
\label{dist_max}
\earr
respectively.

In order to compare these different algorithms, it is useful to
recall the mathematical definition of distance. Given a set $\cS$,
a distance (or a metric) $\delta$ is a non-negative application
\beq d: \cS \times \cS
\longrightarrow \mathbb{R} , \label{distfunc}
\eeq
endowed with the following properties, valid $\forall x,y \in
\cS$:
\barr
& & d(x,y)=0 \quad \Longleftrightarrow \quad x=y, \label{assio_1} \\
& & d(x,y)=d(y,x), \label{assio_2} \\
& & d(x,y)\leq d(x,z)+d(y,z) , \quad \forall z \in \cS .
\label{assio_3}
\earr
Incidentally, notice that symmetry (\ref{assio_2}), as well as
non-negativity, are not independent assumptions, but easily follow
from (\ref{assio_1}) and the triangular
inequality~(\ref{assio_3}).

It is not difficult to prove from the very definition
(\ref{link_haus}) that the Hausdorff distance  between compact and
non-empty sets satisfies (\ref{assio_1})-(\ref{assio_3}). On the
other hand, (\ref{dist_min}) and (\ref{dist_max}) are \emph{not}
distances: the former does not satisfy the triangular inequality
(\ref{assio_3}), while the latter does not fulfil the basic
requirement (\ref{assio_1}), $d_{\rm C}(A,A)\neq0$, for any
compact set containing more than one point: in this sense, it
performs a sort of coarse graining over the data set. The
Haussdorf function, being a distance in a strict mathematical
sense, enables us to rest on sound mathematical ground.

The Hausdorff distance has never been used (to the best of our
knowledge) in the context of clustering. It is a useful tool in
the analysis of complex sets, with complicated (and even
fractal-like) structures. It is in such a case that one expects
that Hausdorff behave better than the other methods, since it
relies on rigorous mathematical concepts.

\section{Application to Financial Data}
\label{sec:financial}

We now apply the Hausdorff linkage algorithm to a topic of growing
interest: the analysis of financial time series. In particular, we
focus on the $N=30$ shares composing the DJIA index, collecting
the daily closure prices of its stocks for a period of 5 years
(1998-2002). We chose this index for two reasons. First, because
these data are easily accessible. The second, and more important
reason is the ``quality" (in the sense of reliability) of prices.
The DJIA index, indeed, aggregates the shares of some of the more
valuable and capitalized world corporations, so that their prices
are highly contributed by market makers. This means that we always
expect to find, even in the worst possible scenario, a financial
intermediator (market maker) ready to quote both bid and offer
prices for these assets. For this reason, these shares are very
frequently traded. In financial terminology, they are said to be
``liquid."

\begin{figure}
\begin{center}
\includegraphics[width=0.7\textwidth]{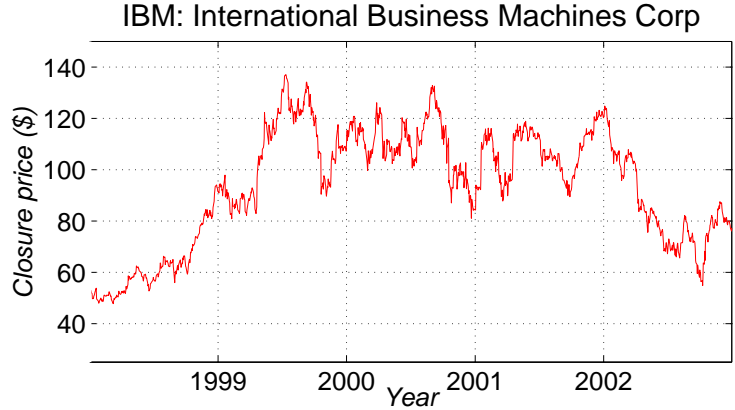}
\end{center}
\caption{Time evolution of the closure price of a stock value
(\textsf{IBM}), for the period 1998-2002.} \label{fig:C_IBM}
\end{figure}

\begin{figure*}
\includegraphics[width=\textwidth]{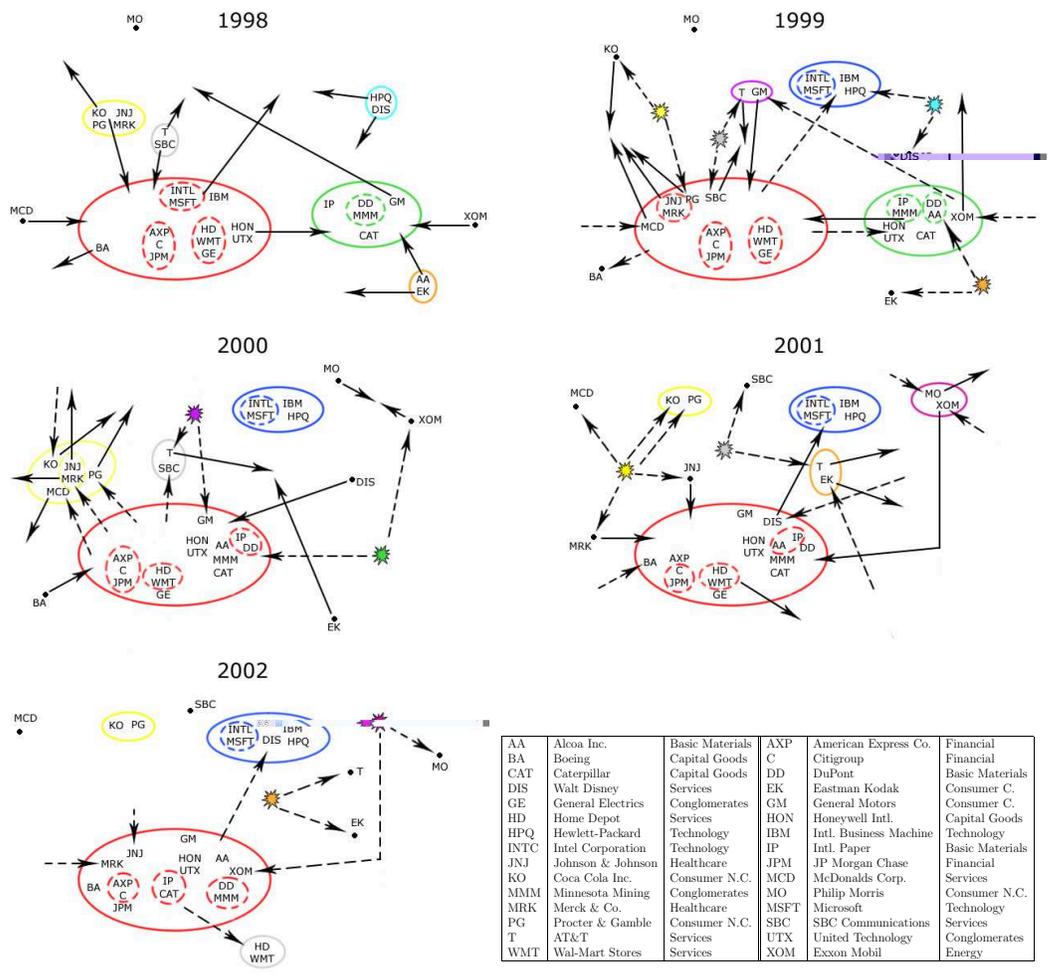}
\caption{Clusters obtained by analyzing the daily logarithm
closure price difference time series during 1998-2002. The
innermost subclusters are indicated with a dashed bubble. Dashed
arrows = past; full arrows = future. The position of the points
representing the stocks is not directly related to the distance
matrix (\ref{dist}) and has no effective ``spatial" meaning: the
pictorial representation simply reflects the aggregation of points
and subclusters into larger clusters. Bottom right: acronyms
(tickers) of the stocks and related industries (C.\ = Cyclical;
N.C.\ = Non-Cyclical; Intl. = International)
}\label{fig:blobs98-02}
\end{figure*}

Figure \ref{fig:C_IBM} displays the typical behavior of a stock
value (\textsf{IBM}) for the investigated time period. The
companies of the DJIA stock market are reported in Figure
\ref{fig:blobs98-02} (bottom right), together with the
corresponding industries.
We will look at the temporal series of the daily logarithm closure
price differences \beq Y_{i}(t) \equiv \ln P_{i}(t)-\ln
P_{i}(t-1),
\label{ln_C} \eeq where $P_{i}(t)$ is the closure price of the $i$th
share at day $t$. Both $P_i$ and $Y_i$ are very irregular
functions of time. In order to quantify the degree of similarity
between two time series and use our linkage algorithm we adopt the
following metric function, that quantifies the synchronicity in
their time evolution \cite{mantegna,mant_stan,Grilli1}
\beq
d_{ij} = \sqrt{2(1-c_{ij})}\label{dist}~,
\eeq
where $c_{ij}$ are the correlation coefficients computed over the
investigated time period:
\beq
c_{ij} = \frac{\langle Y_{i}Y_{j}\rangle - \langle
Y_{i}\rangle\langle Y_{j}\rangle}{\sqrt{(\langle Y_{i}^{2}\rangle
- \langle Y_{i}\rangle^{2})(\langle Y_{j}^{2}\rangle - \langle
Y_{j}\rangle^{2})}}  \label{rho}
\eeq
and the brackets denote the average over the time interval of
interest (one year in our case). Table \ref{tab_corr} displays a
part of the $N\times N$ matrix of the correlation coefficients
(year 1998). It is worth stressing that almost all correlation
coefficients are positive, with values not too close to 1, thus
confirming that, in many cases, stocks belonging to the same
market do not move independently from each other, but rather share
a similar temporal behavior. The distance (\ref{dist}) is a proper
metric in the ``parent" space, ranging from 0 for perfectly
correlated series ($c_{ij} = +1$) to 2 for anticorrelated stocks
($c_{ij} = -1$). (The representative points lie therefore on a
hypersphere.)
\begin{table}
\begin{center}
\begin{tabular}{c|c|c|c|c|c|}
    & ~~~\textsf{AA}~~~ & \textsf{AXP} & \textsf{BA} & \textsf{CAT} & \textsf{C} \\ \hline
  \textsf{AA} & ~~~~1~~~~ & 0.37004 & 0.22458 & 0.3568 & 0.3508 \\ \hline
  \textsf{AXP} &   & 1 & 0.35461 & 0.41916 & 0.61247 \\ \hline
  \textsf{BA} &   &   & 1 & 0.32852 & 0.26917 \\ \hline
  \textsf{CAT} &   &   &   & 1 & 0.33937 \\ \hline
  \textsf{C} &   &   &   &   & 1 \\ \hline
\end{tabular}
\end{center}
\caption{\small A part of the matrix of the correlation
coefficients $c_{ij}$ (\ref{rho}) for the temporal series of the
daily logarithm price differences of the stocks composing the DJIA
index (year 1998). The acronyms (tickers) are explained in Figure
\ref{fig:blobs98-02}. }\label{tab_corr}
\end{table}

\section{Results and Discussion}

Figure \ref{fig:blobs98-02} shows the results of our analysis
based on the Hausdorff ansatz. Rather than showing the
dendrograms, we prefer to give a pictorial representation of the
evolution of the stocks by using bubbles to represent clusters and
arrows to represent the movements of the stocks. Some innermost
subclusters are indicated with a dashed bubble and full (dashed)
arrows denote future (past) movements. A small ``exploding" star
represents a bubble/cluster that disappears.

It is very interesting and challenging to try and analyze, from a
mere economic viewpoint, some of the movements in the graphs, in
order to catch some ``a posteriori" hints about the dynamics of
the stocks. At first sight, one clearly recognizes that some of
the clusters correspond to homogeneous groups of companies
belonging to the same industry: this is the case of the financial
services firms \{\textsf{AXP}, \textsf{JPM}\, \textsf{C}\}, retail
companies \{\textsf{HD}, \textsf{WMT}\}, companies dealing with
basic materials (\textsf{AA}, \textsf{IP}, \textsf{DD}),  the
technological core \{\textsf{IBM}, \textsf{INTC}, \textsf{MSFT},
\textsf{HPQ}\} and the health care firms \{\textsf{JNJ},
\textsf{MRK}\}.

Moreover, one observes a large super-cluster made up of 10-15
stocks (financial, conglomerates, services, capital goods),
containing some homogenous subclusters, which is more or less
stable during the whole 5-year period investigated.

It is worth stressing, between 1998 and 1999, the migration of the
hi-tech companies \{\textsf{IBM}, \textsf{INTC}, \textsf{MSFT}\}
from this cluster. At the end of these two years, they end up
forming a separated cluster with \textsf{HPQ}, that remains stable
for all the following period. As is well known, 1999 is the year
when the high-tech bubble started to grow up. Even more
interesting is the ``path" of Disney. During 1998 it is perceived
to be linked to
\textsf{HP}, which was (and still is) its favorite supplier of
hardware. Then, during the following years, it remains more or
less single, until, between 2001 and 2002, it rejoins \textsf{HP}
into the high-tech core. This evolution can probably be explained
by remembering Disney's strategic efforts to increase its Media
Network segment, that consisted also in a series of acquisitions
(the last two: Fox Family Worldwide Inc.\ and Baby Einstein Co).

We emphasize that these remarks are not an input of our analysis:
our clustering algorithm is purely \emph{mathematical}, and no
genuinely ``economical" information (e.g., on industrial
homogeneity) was used at the outset. In this sense the position
and movements of the stocks in the figures are implied from the
market itself.

The definition of the mutual positioning of companies can have an
immediate pertinence in a matter of great interest for financial
institutions: the portfolio optimization. In a few words (and
without entering into complex matters), portfolio theory suggests
that in order to minimize the risk involved in a financial
investment, one should diversify among different assets by
choosing those stocks whose price time evolutions are as diverse
as possible (it is never safe to put all the eggs into a single
basket). Moreover, this strategy must be continuously updated, by
changing weights and components, in order to follow the market
evolution. In the framework we presented, by investigating the
shares' behavior and tracking the evolution of their mutual
interactions, a first, crude portfolio-optimization rule that
emerges would be: choose stocks belonging to clusters that are as
``distant" as possible from each other.

In conclusion, we have introduced a novel clustering procedure
based on the Hausdorff distance between sets. This genuinely
mathematical method was used to investigate the time evolution of
the stocks  belonging to the DJIA index. We found the resulting
partitions through the 5-year period investigated to be
significant from an economical viewpoint and suited to a
meaningful \emph{a posteriori} analysis and interpretation. We
believe that this technique is able to extract relevant
information from the raw market data and yield meaningful hints
for the investigation of the mutual time evolution of the stocks.
For the same reasons this procedure could be implemented as the
first step towards an evolved portfolio selection and optimization
procedure.

\vspace{.5cm}
\textbf{Acknowledgements.}
We thank Sabrina Diomede for a discussion and a pertinent remark.




\begin{thebibliography}{15}





\bibitem{fukunaga} K. Fukunaga, \textit{Introduction to Statistical Pattern
Recognition} (Academic Press, San Diego, 1990).

\bibitem{jain} A. K. Jain and R. C. Dubes, \textit{Algorithms for Clustering
Data} (Prentice Hall, New York, 1988).

\bibitem{central} A. Gersho and R. M. Gray, \textit{Vector Quantization and
Signal Processing} (Kluwer Academic Publisher, Boston, 1992).

\bibitem{duda} R. O. Duda, P. E. Hart and D. G. Stork, \textit{Pattern
Classification} (John Wiley \& Sons, New York, 2002).

\bibitem{hofmann} T. Hofmann and J. M. Buhmann, \textit{Pairwise Data Clustering by
Deterministic Annealing}, IEEE Transaction on Pattern Analysis and
Machine Intelligence, \textbf{19}, 1 (1997).

\bibitem{edgar} F. Hausdorff, \textit{Grundz\"{u}ge der
Mengenlehre} (von Veit, Leipzig,  1914). [Republished as
\textit{Set Theory}, 5th ed. (Chelsea, New York, 2001).]

\bibitem{mantegna} R. N. Mantegna, Eur. Phys. J. B \textbf{11}, 193 (1999).

\bibitem{mant_stan} R. N. Mantegna and H. E. Stanley, \textit{Introduction to
Econophysics} (Cambridge University Press, 2000).

\bibitem{Grilli1}
M. Bernaschi, L. Grilli and D. Vergni, Physica A \textbf{308}, 381
(2002); L. Grilli, Physica A \textbf{332}, 441 (2004).

\end{thebibliography}
\end{document}